\documentclass[12pt]{article}
\input epsf.sty
\newcommand{\nl}{\hspace{-.65cm}}
\newcommand{\be}{\begin{equation}}
\newcommand{\ee}{\end{equation}}
\newcommand{\ben}{\begin{eqnarray}\displaystyle}
\newcommand{\een}{\end{eqnarray}}

\usepackage{amssymb}
\usepackage{amsmath}
\usepackage{comment}
\usepackage{mathrsfs}
\usepackage[all]{xy}
\usepackage[dvips]{graphicx}
\usepackage{epsfig,color}
\def\be{\begin{equation}}
\def\ee{\end{equation}}
\def\ba{\begin{align}}
\def\ea{\end{align}}

\begin{document}

\begin{center}
{
\Large{String Theory and  The Arrow of Time
 }
}
\vspace{6mm}

 { Nissan Itzhaki}
\vspace{6mm}
\break
\textit{
 Tel Aviv University, Ramat Aviv, 69978, Israel
}

\end{center}

\vspace{2mm}

\begin{abstract}

Time-like linear dilaton triggers, at the classical level,  the creation of  closed folded strings at an instant. We show that in cosmology
 these instant folded strings  induce  negative pressure at no  energy cost.
Hence they seem to allow an era in which the energy density increases (decreases) while the universe is  expanding (contracting). This and other aspects of instant folded strings suggest that they might shed new light on the origin of the arrow of time.

\end{abstract}

\newpage

\baselineskip=18pt


Despite the remarkable progress \cite{Spergel:2006hy,Akrami:2018odb} made in making contact between observation and inflation \cite{Guth:1980zm}, cosmology is still full of mystery.
In particular, the origin of the arrow of time is still not clear  (see e.g. \cite{Harlow:2011az,Carroll:2014uoa} and references therein).
In the early universe, it is related to the initial condition problem: how come the universe begins at a state with such low entropy? In late times, the   Boltzmann brains  problem was resurrected by the cosmological constant \cite{Dyson:2002pf}. Moreover,  the  cosmological constant problem could be related to these issues \cite{Steinhardt:2006bf,Bousso:2011aa}.

One of the things that makes it difficult to address  these and related questions is the fact that normally the
energy density, $\rho$, cannot decrease (increase)  as the universe contract (expands). This follows from the
 first law of thermodynamics
\be
\dot{\rho}=-3H(\rho+p),
\ee
and the Null Energy Condition (NEC)
\be\label{nec} \rho +p\geq 0,\ee
where $p$ is the pressure.

Excitations that violate (\ref{nec}) are interesting from this perspective.
There are  exotic models that violate (\ref{nec})  (for a review see  \cite{Rubakov:2014jja}), but it is not clear if they
 can be realized in string theory.
Exotic excitations that do appear in string theory are  tachyons that even  in the  closed strings sector are often under control \cite{Adams:2001sv}. However tachyonic excitations do not violate (\ref{nec}).\footnote{There are  objects with negative energy density  in string theory - orientifolds \cite{Sagnotti:1987tw,Horava:1989vt,Gimon:1996rq}. These, are excluded from the present discussion since they  modify the boundary condition.}

In this short note we wish to  describe excitations in string theory that do  violate (\ref{nec}).
We hope that a careful study of these excitations could shed light on these conceptual questions.

In   \cite{Itzhaki:2018glf}  it was shown that in  certain situations, that involve a time-like linear dilaton, closed folded  strings can appear {\it classically} at an instant. We refer to such strings as Instant Folded Strings (IFS). Since  created classically at an instant, energy conservation implies that the total energy  of an IFS must vanish \cite{Attali:2018goq}.
 Consequently  they do not contribute to   $\rho$ when distributed homogeneously, e.g. in cosmology. Like any other string, an IFS  does admit   a nontrivial energy momentum tensor. Hence   they do generate pressure in cosmology.  Below we show that this pressure is negative which implies
 that, in a sense, the IFSs violate (\ref{nec})  maximally - they generate negative pressure at no cost of  energy.

We start by reviewing the IFS. Consider  string theory on a 2D background, that involves  a time-like linear dilaton
\be\label{bac}
ds^2=-dt^2+dx^2,~~~\Phi=\Phi_0+Q t,~~~\mbox{with}~~~~Q>0
\ee
times a manifold ${\cal M}$. The interesting physics takes place in the 2D background and the sole role of ${\cal M}$ here is to ensure that the theory is critical.
This background was studied extensively over the years as a toy model for stringy cosmology (see e.g. \cite{Hellerman:2006nx,Aharony:2006ra} and references therein).

The background (\ref{bac}) admits a curious classical string solution, that is  obtained,   via an analytic continuation, from a solution found by Maldacena  some time ago  in the case of space-like linear dilaton \cite{Maldacena:2005hi}. The solution takes the form  \cite{Itzhaki:2018glf}
\be\label{s}
x=x_0+\sigma,~~~~t=t_0+Q \log\left( \frac12\left( \cosh\left(\frac{\sigma}{Q}\right)+\cosh\left(\frac{\tau}{Q}\right)\right)\right),
\ee
with $-\infty < \sigma, \tau < \infty$. For $t<t_0$ there is no string at all. $t=t_0$ when $\tau=\sigma=0$. For $t>t_0$ the string configuration is symmetric under   $\tau \to -\tau$.  The solution describes a closed folded string, with a fold at  $\tau=0$, that is spontaneously created at $x_0$ and $t_0$.
The  string  is created  at zero size  and it grows faster than light. 
 In a sense, one can view an IFS as a two dimensional stringy baby universe.\footnote{ At least naively, U-duality seems to  suggest  also instant Dp-branes solutions that can be viewed as (p+1)-dimensional baby universes. For example, acting with S-duality on an IFS might give an instant D1-brane in the background (\ref{bac}) with $Q<0$. This is not certain since S-duality acts in a definite way only on BPS states. If exist, instant Dp-branes are not expected to be found at the DBI level - analysis at the boundary state level is required.  This follows from the fact that the $\alpha^{'}$ corrections to the Virasoro constraints are crucial for \cite{Maldacena:2005hi}. }

The relation between the IFS and the arrow of time is threefold.  First, the fact that (\ref{s}) describes a string that is created at an instant leads to the  unusual feature that the arrow of time, as defined on the IFS,  flips sign.
There are two natural ways to define an arrow of time  on an IFS. The first, which is valid for every string, is  
\be 
A_1=\frac{dt}{d\tau} .
\ee 
In the case of a folded string, one can also treat the fold as a particle and define the arrow of time accordingly:
\be 
A_2=\frac{dt}{d\sigma}~~~ \mbox{at} ~~~ \tau=0.
\ee 
In both definitions the arrow of time flips sign (see figure 1). The time scale associated with this  flipping is of the order of $ Q$.
Our main focus is on  $Q\ll1$. In that case, the flipping time scale is much shorter than the string scale and the scale set by the background (\ref{bac}), $1/Q$.

Second, the IFS appears to link the time-like  gradient of the dilaton with the thermodynamic arrow of time. Acting with time reversal on (\ref{bac}) takes $Q\to -Q$. The  evolution forward in time with $Q>0$ is drastically different than that with $Q<0$, even when $|Q|\ll 1$.    Negative $Q$ does not allow IFS creation and the IFS vacuum evolves into the vacuum. Namely, if at $t=-\infty$ there are no IFSs then there will be none also at any finite $t$. However, for positive $Q$ the situation is very different. Even if we impose that at $t=-\infty$ there are no IFSs, the solution (\ref{s}) means that they will be created rapidly at later times.   The amplitude   to find at $t_1$  a closed folded string that is stretched from $x_1$ to $x_2$, subject to the initial condition that there are no IFSs at $t\to-\infty$ is given, to leading order, by $\exp(i S)$ where $S$ is the action associated with the solution (\ref{s}) with the relevant boundary condition.  Since $S$ is real   the production rate is not  exponentially suppressed  and so (\ref{bac}), with $Q>0$, gets filled with IFSs.  Hence only for positive $Q$ the entropy, associated with the IFSs, grows with time.

In short, for $Q>0$ ($Q<0$) IFSs can be created (annihilated) classically, but they  cannot be 
annihilated (created) classically. As a result, the thermodynamic arrow of time, which in the present case is determined by the number of IFSs, is fixed by $Q$. 
Note that the standard problem with the  thermodynamic arrow of time - that according to fundamental physics the entropy should increase also when going backward in time - does not appear here since an IFS is created at an instant. Going backward in time (with positive $Q$)  we find fewer and fewer IFSs, which leads to a smaller and smaller  entropy. 

The third relation to the arrow of time is that, as we now show,  IFSs violate the NEC in cosmology.
 The energy-momentum associated with (\ref{s}) was calculated in \cite{Attali:2018goq}. For simplicity, we present here the result  for $Q\ll1$, but the conclusions that follow are general. For $Q\ll1$  the energy-momentum tensor associated with (\ref{s}) takes a particularly simple form
\be\label{lo}
T_{uu}=-\frac{v}{2\pi\alpha^{'}} \delta(u) \theta(v), ~~~T_{vv}=-\frac{u}{2\pi\alpha^{'}}\delta(v)\theta(u),~~~T_{uv}=\frac{1}{2\pi\alpha^{'}}\theta(u)\theta(v),
\ee
with  $u=(t-t_0)+(x-x_0)$ and $v=(t-t_0)-(x-x_0)$. The minus sign in $T_{uu}$ and $T_{vv}$ reflects the fact  that there is a {\it negative} null flux at the folds of the string. This flux  becomes more negative with time in order to compensate for the positive energy stored in the bulk of the IFS. As a result, a single IFS violates the Averaged NEC \cite{Attali:2018goq}, and, as we now show, IFSs generate  negative pressure in cosmology.

Since the background (\ref{bac}) is invariant under translation in $x$ we expect
the density of IFSs that are  created at $(t_0, x_0)$ to depend on $t_0$, but not on $x_0$. Denoting this density by $\rho_{IFS}(t_0)$ we find, by integrating (\ref{lo}),  that the only component of the energy-momentum tensor that does not vanish is
\be\label{pres}
T_{xx}(t)=- \frac{1}{2\pi\alpha^{'}} \int ^{t} dt_0 \rho_{IFS}(t_0) (t-t_0).
\ee
Namely,  the IFSs induce negative pressure at no cost of energy.
Eq.  (\ref{lo}) holds for any positive $Q$ (not necessarily small) as long as $t-t_0\gg Q$. Hence, for any $Q$,    (\ref{pres}) becomes,  very quickly, a good approximation.

\begin{figure}
\begin{center}
\includegraphics[scale=0.45]{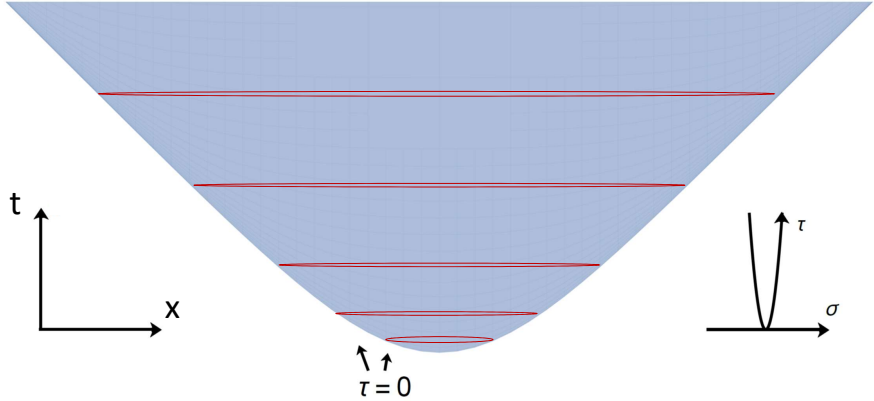}\vspace{-25mm}
\caption{The instant folded string configuration. Defining the arrow of time as $dt/d\tau$ we see that it flips sign at $\tau=0$. One might argue that a more natural definition of the arrow of time, that compares well with a point like trajectory,  is $dt/d\sigma$ at the fold ($\tau=0$). This definition too leads to an arrow of time flipping at $\sigma=0$.   The red squeezed rings illustrate that this configuration describes a closed folded string that is created at an instant and grows rapidly with time. }
\label{SBHpen}
\end{center}
\end{figure}\label{fh1}

Some comments:

\nl $\bullet$  As discussed above since the IFSs are created classically we do not expect $\rho_{IFS}(t_0)$ to be exponentially suppressed. Such a suppression is usually  attributed to tunneling. Eq. (\ref{s}) is, however,  a solution in Lorentzian signature. In fact,  in the appendix we estimate that  $\rho_{IFS}(t_0)\sim 1/\alpha^{'}$. 
This combined with 
 the  $t-t_0$ in (\ref{pres}) implies that the negative pressure grows rapidly once IFSs are created.

\nl $\bullet$ In 4D time-like dilaton background 
\be\label{four}
ds^2=-dt^2+dx^2+dy^2+dz^2,~~~\Phi=\Phi_0+Q t,
\ee
we reach  similar conclusions. Namely, IFSs are created classically with the same solution (\ref{s}), only that now  they can stretched in any spatial direction. The background (\ref{four}) is invariant under translation and rotation. Hence the IFSs are distributed  homogeneously and
isotropically, which again leads to  negative pressure at no energy cost.

\nl $\bullet$ The length scale associated with an IFS creation is small - of the order of $Q$. Therefore we expect IFSs to be created whenever the local inertial frame takes the form of (\ref{four}) and their density, $\rho_{IFS}$, to be determined locally. This implies that
(\ref{pres}) is a good approximation as long as $t-t_0$ is smaller than the length scale set by the curvature or the second derivative of the dilaton.

These comments suggest  that we might want to consider the following   scenario.  Suppose that in the early universe there was a small region of space-time  that is described by  (\ref{four}).  In this region IFSs will be created. The factor of $(t-t_0)$  in (\ref{pres}) implies that they will generate quickly a large negative pressure at no cost of energy. Consequently, this region will expand rapidly. At this stage several things are expected to happen. Due to the acceleration of the universe light degrees of freedom will be excited -  the universe  will warm up. Simultaneously, the impact of the IFSs will get smaller for several reasons. First, as discussed above,  (\ref{pres}) is valid only for distances smaller than the curvature.  Indeed causality implies that once  the size of the IFS exceeds the Hubble scale their backreaction  will be suppressed. Second, interactions with standard matter and among themselves should lead to IFSs annihilation. Third, as was shown in \cite{Giveon:2020xxh} the backreaction of the IFSs tends to decrease $Q$ which  triggers their creation.

This  suggests that after a short exotic period of IFSs domination, in which the NEC is violated and $\rho$ increases as the universe expands rapidly,
the universe will enter a standard radiation dominated era.  
So in this setup a positive $Q$ is  the spark that ignites the big-bang.

This is just one possibility. More generally, objects that violate the NEC and are triggered by time-like linear dilaton could play a key role in cosmology. For example they can be helpful   in some fascinating models of cosmology (e.g. \cite{Gasperini:1992em,Brustein:1994kw,Khoury:2001wf,Khoury:2001bz}) that are in tension with classical theorems in general relativity that assume the NEC.

Another topic that might be interesting to revisit is  the possibility of creating, in principle,  a universe in the lab. In \cite{Farhi:1986ty} it  was shown that assuming the NEC in the context of general relativity this too is ruled out by Penrose singularity theorem \cite{Penrose:1964wq}. The fact that IFSs violate  the NEC suggests that it might be interesting to reconsider this in  the context of string theory.

A  rigorous examination of these possibilities involves a detailed calculation of the creation and annihilation rates of the IFSs.

\vspace{10mm}

\section*{Acknowledgments}
We thank R. Brustein, A. Giveon, A. Hashimoto,  U. Peleg and J. Troost for discussions.
This work is supported in part by a center of excellence supported by the ISF (grant number 2289/18) and BSF (grant number 2018068).

\vspace{10mm}



\begin{appendix}

\section{ An estimation of $\rho_{IFS}$}

In this appendix we estimate $\rho_{IFS}$ in the background (\ref{bac}) using the uncertainty principle and (\ref{lo}).

Suppose that at $t\to -\infty$ there are no IFS. To determine   $\rho_{IFS}(t_0, x_0)$ we ask what is the probability to find, with this initial condition, an IFS that is created at $(t_0, x_0)$. Alternatively,
we can ask what is the probability to find an IFS at $t_1>t_0$ that folds at $x_1$ and $x_2$. For $t_1-t_0\gg Q$ the two questions are identical if $x_1=x_0+(t_1-t_0)$ and $x_2=x_0-(t_1-t_0)$ (see figure 2).
For every $x_1, x_2$ and $t_1$ there is a single classical configuration that fits these boundary condition. Since the classical configuration is Lorentzian the amplitude $A(t_1, x_1, x_2)\sim \exp(iS)$ and so $P(t_1, x_1, x_2)=|A(t_1, x_1, x_2)|^2\sim 1$.
Since $x_1$ and $x_2$ are fixed completely at  $t_1$ a finite $P$ implies that $\rho_{IFS}=\infty$, which, of course, makes no sense.

Quantum effects are expected to render $\rho_{IFS}$ finite. A simple way to see this is the following. Suppose that we wish to measure where the folded string was created. 
To determine $v_0=t_0+x_0$ we have to measure $x_2$ at $t_1$.
To determine $u_0=t_0-x_0$ we have to measure $x_1$ at $t_1$. However, the two measurements do not commute. To see this  note that  (\ref{lo}) implies a relation between $u_0$ and the null momentum $ P_v =\int dv T_{vv} (U)$ (marked by the left green arrow in figure 2)
\be
u_0=u_1+ 2\pi \alpha^{'} P_v.
\ee
The uncertainty principle, $\Delta v \Delta P_v>1,$ gives
\be\label{hg}
\Delta u_0 \Delta v_0 \geq \alpha^{'},
\ee
which is equivalent to saying  that at $t_2$
\be\label{hj}
\Delta x_1 \Delta x_2 \geq \alpha^{'}.
\ee
We conclude, using either (\ref{hg}) or (\ref{hj}),  that
\be
\rho_{IFS}\sim 1/\alpha^{'}.
\ee

\begin{figure}
\begin{center}
\includegraphics[scale=0.45]{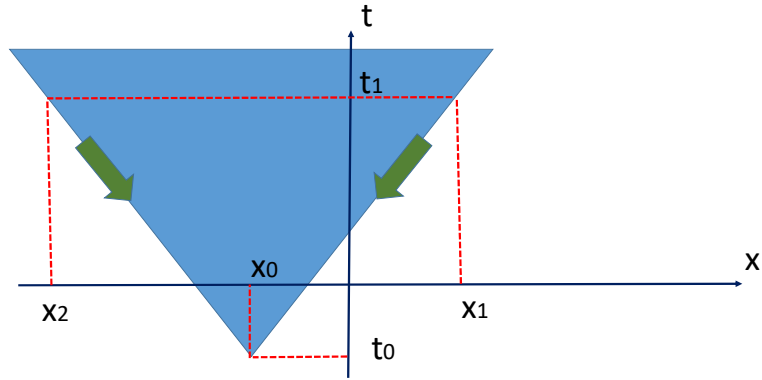}\vspace{-30mm}
\caption{For every $x_1,$ and $x_2$ at $t_1$ there is a classical IFS configuration that did not exist at $t\to-\infty.$}
\end{center}
\end{figure}

\end{appendix}



\bigskip




\begin{thebibliography}{24}


\bibitem{Spergel:2006hy}
D.~N.~Spergel \textit{et al.} [WMAP],
``Wilkinson Microwave Anisotropy Probe (WMAP) three year results: implications for cosmology,''
Astrophys. J. Suppl. \textbf{170}, 377 (2007)
doi:10.1086/513700
[arXiv:astro-ph/0603449 [astro-ph]].




\bibitem{Akrami:2018odb}
Y.~Akrami \textit{et al.} [Planck],
``Planck 2018 results. X. Constraints on inflation,''
Astron. Astrophys. \textbf{641}, A10 (2020)
doi:10.1051/0004-6361/201833887
[arXiv:1807.06211 [astro-ph.CO]].


\bibitem{Guth:1980zm}
A.~H.~Guth,
``The Inflationary Universe: A Possible Solution to the Horizon and Flatness Problems,''
Adv. Ser. Astrophys. Cosmol. \textbf{3}, 139-148 (1987)
doi:10.1103/PhysRevD.23.347

\bibitem{Harlow:2011az}
D.~Harlow, S.~H.~Shenker, D.~Stanford and L.~Susskind,
``Tree-like structure of eternal inflation: A solvable model,''
Phys. Rev. D \textbf{85}, 063516 (2012)
doi:10.1103/PhysRevD.85.063516
[arXiv:1110.0496 [hep-th]].




\bibitem{Carroll:2014uoa}
S.~M.~Carroll,
``In What Sense Is the Early Universe Fine-Tuned?,''
[arXiv:1406.3057 [astro-ph.CO]].


\bibitem{Dyson:2002pf}
L.~Dyson, M.~Kleban and L.~Susskind,
``Disturbing implications of a cosmological constant,''
JHEP \textbf{10}, 011 (2002)
doi:10.1088/1126-6708/2002/10/011
[arXiv:hep-th/0208013 [hep-th]].

\bibitem{Steinhardt:2006bf}
P.~J.~Steinhardt and N.~Turok,
``Why the cosmological constant is small and positive,''
Science \textbf{312}, 1180-1182 (2006)
doi:10.1126/science.1126231
[arXiv:astro-ph/0605173 [astro-ph]].



\bibitem{Bousso:2011aa}
R.~Bousso,
``Vacuum Structure and the Arrow of Time,''
Phys. Rev. D \textbf{86}, 123509 (2012)
doi:10.1103/PhysRevD.86.123509
[arXiv:1112.3341 [hep-th]].



\bibitem{Rubakov:2014jja}
V.~A.~Rubakov,
``The Null Energy Condition and its violation,''
Usp. Fiz. Nauk \textbf{184}, no.2, 137-152 (2014)
doi:10.3367/UFNe.0184.201402b.0137
[arXiv:1401.4024 [hep-th]].


\bibitem{Adams:2001sv}
A.~Adams, J.~Polchinski and E.~Silverstein,
``Don't panic! Closed string tachyons in ALE space-times,''
JHEP \textbf{10}, 029 (2001)
doi:10.1088/1126-6708/2001/10/029
[arXiv:hep-th/0108075 [hep-th]].

\bibitem{Sagnotti:1987tw}
A.~Sagnotti,
``Open Strings and their Symmetry Groups,''
[arXiv:hep-th/0208020 [hep-th]].


\bibitem{Horava:1989vt}
P.~Horava,
``Strings on World Sheet Orbifolds,''
Nucl. Phys. B \textbf{327}, 461-484 (1989)
doi:10.1016/0550-3213(89)90279-4


\bibitem{Gimon:1996rq}
E.~G.~Gimon and J.~Polchinski,
``Consistency conditions for orientifolds and d manifolds,''
Phys. Rev. D \textbf{54}, 1667-1676 (1996)
doi:10.1103/PhysRevD.54.1667
[arXiv:hep-th/9601038 [hep-th]].

\bibitem{Itzhaki:2018glf}
  N.~Itzhaki,
  ``Stringy instability inside the black hole,''
  JHEP {\bf 1810}, 145 (2018)
  doi:10.1007/JHEP10(2018)145
  [arXiv:1808.02259 [hep-th]].



\bibitem{Attali:2018goq}
K.~Attali and N.~Itzhaki,
``The Averaged Null Energy Condition and the Black Hole Interior in String Theory,''
Nucl. Phys. B \textbf{943}, 114631 (2019)
doi:10.1016/j.nuclphysb.2019.114631
[arXiv:1811.12117 [hep-th]].



\bibitem{Hellerman:2006nx}
  S.~Hellerman and I.~Swanson,
  ``Cosmological solutions of supercritical string theory,''
  Phys.\ Rev.\ D {\bf 77}, 126011 (2008)
  doi:10.1103/PhysRevD.77.126011
  [hep-th/0611317].

\bibitem{Aharony:2006ra}
  O.~Aharony and E.~Silverstein,
  ``Supercritical stability, transitions and (pseudo)tachyons,''
  Phys.\ Rev.\ D {\bf 75}, 046003 (2007)
  doi:10.1103/PhysRevD.75.046003
  [hep-th/0612031].

\bibitem{Maldacena:2005hi}
  J.~M.~Maldacena,
  ``Long strings in two dimensional string theory and non-singlets in the matrix model,''
  JHEP {\bf 0509}, 078 (2005)
  [Int.\ J.\ Geom.\ Meth.\ Mod.\ Phys.\  {\bf 3}, 1 (2006)]
  doi:10.1088/1126-6708/2005/09/078, 10.1142/S0219887806001053
  [hep-th/0503112].





\bibitem{Giveon:2020xxh}
A.~Giveon, N.~Itzhaki and U.~Peleg,
``Instant Folded Strings and Black Fivebranes,''
JHEP \textbf{08}, 020 (2020)
doi:10.1007/JHEP08(2020)020
[arXiv:2004.06143 [hep-th]].

\bibitem{Gasperini:1992em}
M.~Gasperini and G.~Veneziano,
``Pre - big bang in string cosmology,''
Astropart. Phys. \textbf{1}, 317-339 (1993)
doi:10.1016/0927-6505(93)90017-8
[arXiv:hep-th/9211021 [hep-th]].


\bibitem{Brustein:1994kw}
R.~Brustein and G.~Veneziano,
``The Graceful exit problem in string cosmology,''
Phys. Lett. B \textbf{329}, 429-434 (1994)
doi:10.1016/0370-2693(94)91086-3
[arXiv:hep-th/9403060 [hep-th]].


\bibitem{Khoury:2001wf}
J.~Khoury, B.~A.~Ovrut, P.~J.~Steinhardt and N.~Turok,
``The Ekpyrotic universe: Colliding branes and the origin of the hot big bang,''
Phys. Rev. D \textbf{64}, 123522 (2001)
doi:10.1103/PhysRevD.64.123522
[arXiv:hep-th/0103239 [hep-th]].

\bibitem{Khoury:2001bz}
J.~Khoury, B.~A.~Ovrut, N.~Seiberg, P.~J.~Steinhardt and N.~Turok,
``From big crunch to big bang,''
Phys. Rev. D \textbf{65}, 086007 (2002)
doi:10.1103/PhysRevD.65.086007
[arXiv:hep-th/0108187 [hep-th]].




\bibitem{Farhi:1986ty}
E.~Farhi and A.~H.~Guth,
``An Obstacle to Creating a Universe in the Laboratory,''
Phys. Lett. B \textbf{183}, 149-155 (1987)
doi:10.1016/0370-2693(87)90429-1



\bibitem{Penrose:1964wq}
R.~Penrose,
``Gravitational collapse and space-time singularities,''
Phys. Rev. Lett. \textbf{14}, 57-59 (1965)
doi:10.1103/PhysRevLett.14.57






\end{thebibliography}
\end{document}